\documentclass[5p]{elsarticle}

\usepackage{hyperref}

\journal{Acta Materialia}

\begin{document}

\begin{frontmatter}
\newcommand{\titlestring}{Role of twin and anti-phase defects in MnAl permanent magnets.} 
\title{\titlestring}

\author{Simon Bance}
\fntext[myfootnote]{A significant amount of this work was performed while at; Department of Technology, St P{\"o}lten University of Applied Sciences, St P{\"o}lten, Austria} 
\address{Seagate Technology, 1 Disc Drive, Springtown, Derry, BT48 0BF, Northern Ireland.}
\cortext[mycorrespondingauthor]{Corresponding author}
\ead{simon.bance@seagate.com}

\author{Florian Bittner}
\author{Thomas G. Woodcock}
\author{Ludwig Schultz}
\address{IFW Dresden, Institute for Metallic Materials, Helmholtzstra{\ss}e 20, 01069 Dresden, Germany.}

\author{Thomas Schrefl}
\address{Center for Integrated Sensor Systems, Danube University Krems, Viktor Kaplan Str. 2E, 2700 Wiener Neustadt, Austria.}

\begin{abstract}
We quantify and explain the effects of both anti-phase boundaries and twin defects in as-transformed $\tau$-MnAl by carrying out micromagnetic simulations based closely on the results of microstructural characterization. 
We demonstrate that magnetic domain walls nucleate readily at anti-phase boundaries and are strongly pinned by them, due to anti-ferromagnetic coupling. 
Likewise, twin boundaries reduce the external field required to nucleate domain walls and provide strong pinning potentials, with the pinning strength dependent on the twinning angle. The relative strengths of the known twin defect types are quantified based on the anisotropy angles across their boundaries. 
Samples that have undergone heat treatment are imaged using electron back-scatter diffraction. 
The precise crystallographic orientation is mapped spatially and converted into a number of realistic finite element models, which are used to compute the effects of large concentrations of twin domains in a realistic morphology. 
This is shown to have a negative effect on the remanence coercivity and squareness. The maximum energy product $(BH)_{\mathrm{max}}$ is therefore significantly lower than the theoretical limit of the material and much lower than MnAl permanent magnets that have been further processed to remove twin defects. 
The knowledge gained in this study will allow the optimization of processing routes in order to develop permanent magnets with enhanced magnetic properties.  
\end{abstract}

\begin{keyword}
\texttt{elsarticle.cls}\sep Permanent magnets\sep twin defects \sep anti-phase boundaries
\MSC[2010] 82D40
\end{keyword}
\end{frontmatter}

\section{\label{sec:introduction}Introduction}

In the recent search for new rare-earth-free hard magnetic phases MnAl has gained particular attention due to the low cost and high abundance of the raw materials required.\cite{Coey2014, Mcguiness2015}
While not able to compete with the incredible performance of sintered Nd-Fe-B permanent magnets, which can have high maximum energy products above $(BH)_{\mathrm{max}}=400$ kJ/m$^{3}$ (50 MGOe), MnAl, which has an estimated upper limit of $(BH)_{max} \approx 100$~kJ/m$^3$, could form the basis of a low cost class of general purpose permanent magnets with good machineability, to compete with polymer bonded Nd-Fe-B or ferrites. \cite{Lewis2012, Coey2012}

\begin{figure}
\centering
\includegraphics[width=0.82\columnwidth]{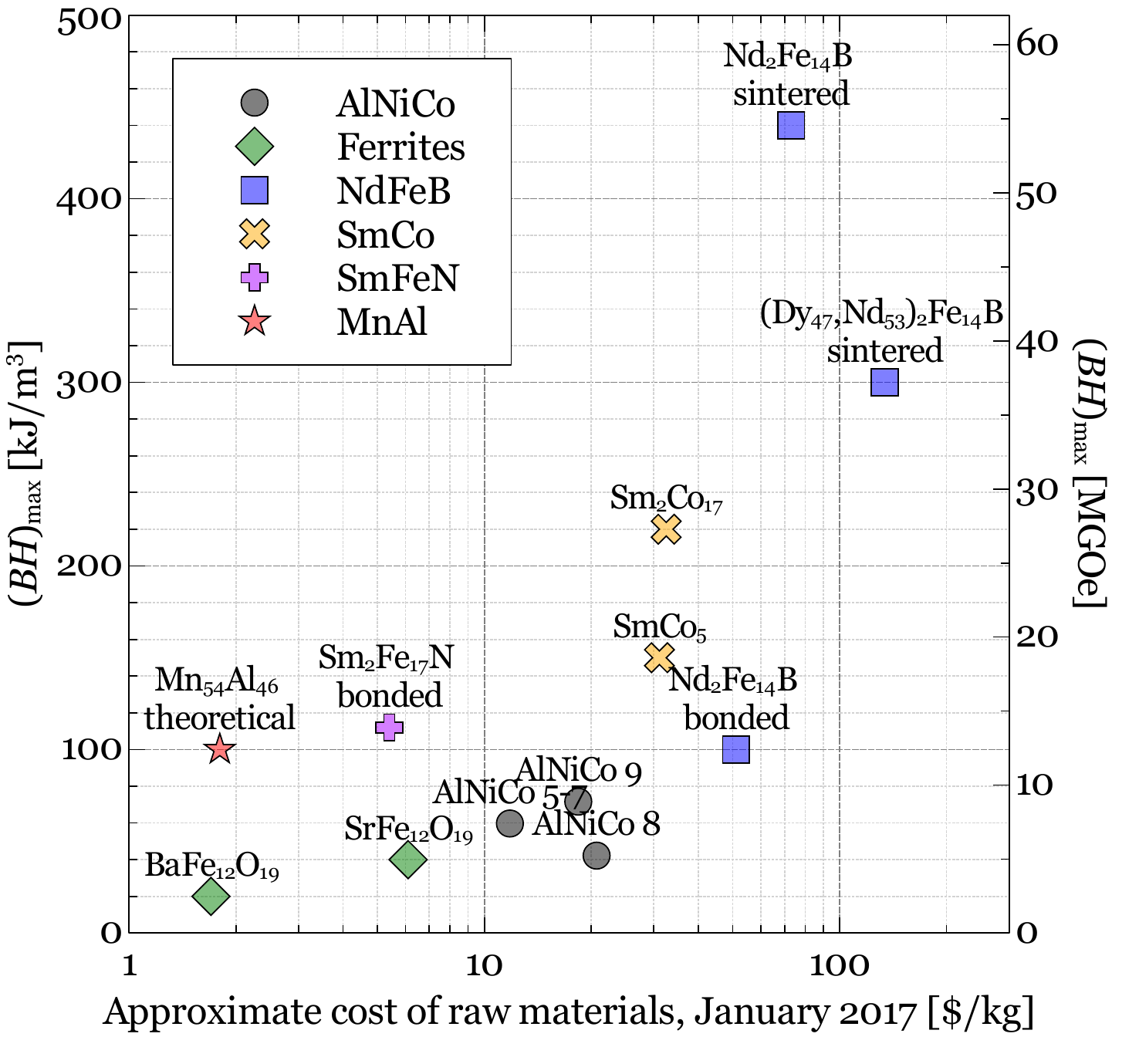}
\caption{Room temperature $(BH)_{\mathrm{max}}$ as a function of approximate raw material costs for the theoretical MnAl permanent magnet and experimental values for a selection of common commercial permanent magnets. The raw material costs are a good indication of the relative cost of the manufactured magnets. }
\label{fig:cost}
\end{figure}  

A plot of approximate $(BH)_{\mathrm{max}}$ versus raw material price in \$/kg for various permanent magnet products is given in Figure \ref{fig:cost}. \footnote{Prices are estimated from the volume-fraction-adjusted market prices of the pure constituent materials alone (without considering any other manufacturing costs) except for the ferrites BaFe$_{12}$O$_{19}$ and SrFe$_{12}$O$_{19}$ where the current market prices for the required amount of precursors, BaCO$_{3}$ and SrCO$_{3}$ are used. }
Typical $(BH)_{\mathrm{max}}$ values and information on composition are collected from the existing literature and from manufacturer data, depending on availability. \cite{Coey2010, gutfleisch2011magnetic, wang2015effect, eclipse, Zhou2014, magnetic2000standard} 
It should be noted that although Sm-based magnets turn out to be quite competitive as mid-performing permanent magnets, the critical supply status of Sm (as with all rare earths) means that the price is likely to increase. 

The proposed MnAl magnets consist of grains of the L1$_{0}$-structured $\tau$ phase, which is characterized by the presence of crystalline defects such as twins,\cite{Jakubovics1977b, VanLanduyt1978a, Yanar2001c} of which there are three different types \cite{Bittner2015a} and anti-phase boundaries (APBs).\cite{Zijlstra1966b, zijlstra_coping_1979} The microstructure is highly sensitive to the metallurgical state of the material. APBs are observed after the phase formation but are absent in hot deformed magnets \cite{VanLanduyt1978a} and the population of 2 of the 3 different twin-like defects is remarkably reduced in the hot deformed state.\cite{Bittner2015a}

In the initial material, after the formation of the $\tau$ phase the coercivity is comparably low, which has been attributed to the presence of APBs.\cite{Houseman1983e} APBs may be responsible for easy nucleation of domain walls but also act as domain wall (DW) pinning potentials, due to a local reduction of the domain wall energy across an APB.\cite{zijlstra_coping_1979} APBs are created by a shear of the lattice of 1/2 [101] and this leads to a shorter atomic spacing between Mn atoms at the boundary, which may give rise to anti-ferromagnetic (AFM) coupling. This divides the crystal into two anti-parallel ferromagnetic regions.

Furthermore the impact of the complicated twinned structure on the magnetic properties may act as pinning centers for domain walls.\cite{Jakubovics1977b, Thielsch2017} 
The orientation of the easy axis changes when crossing the twin boundary and the misorientation angle depends on the type of twin boundary. The strength of pinning interaction of the 3 different types of twins with magnetic domain walls may therefore be different and until now have not been determined. Knowledge of this is a vital step in the development of rare-earth free MnAl magnets. 

The aim of the current work is to understand and quantify the interaction of APBs and twins with domain walls in as-transformed $\tau$-MnAl by carrying out micromagnetic simulations based closely on the results of microstructural characterization. The knowledge gained in this study will allow the optimization of processing routes in order to develop permanent magnets with enhanced magnetic properties.

\section{\label{sec:experiment}Experimental work}

A binary Mn-Al alloy with the nominal composition Mn$_{54}$Al$_{46}$ (at.\%) was prepared by arc melting 99.99\% pure Mn and Al under Ar atmosphere. The composition of the ingot was checked using chemical analysis and it was found that the difference between nominal and actual composition was below 1 at.\%. 

As the cast material consists of a mixture of $\tau$ and equilibrium $\gamma_2$ phases, a homogenization heat treatment at 1100$^\circ$C for 5 days followed by quenching was carried out. For this, the as-cast material was encapsulated in a glass tube which was evacuated to 10$^{-4}$ mbar and then filled with 150 mbar of pure Ar. 

\begin{figure}
\centering
\includegraphics[scale=1]{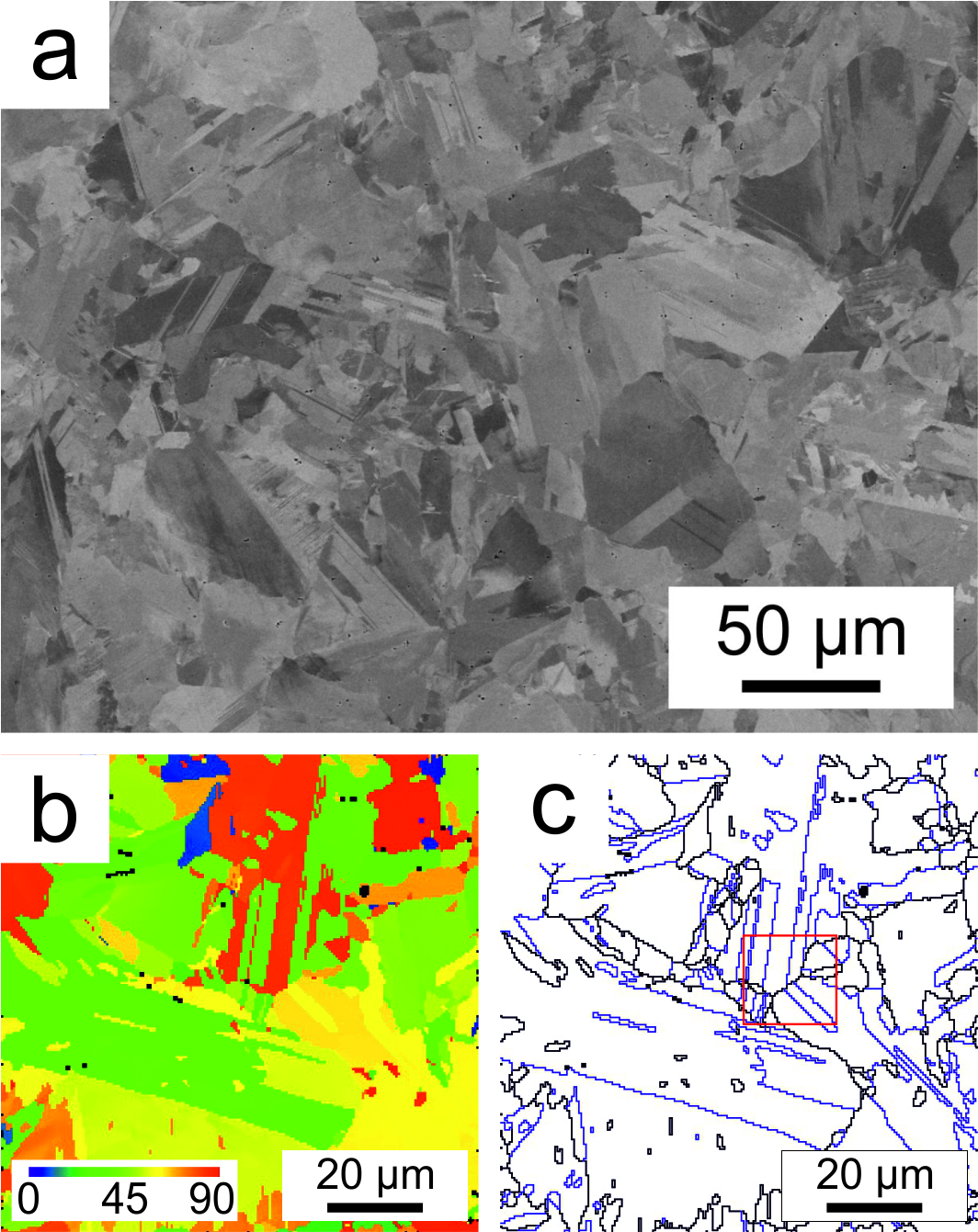}
\caption{Microstructure of Mn$_{54}$Al$_{46}$ after homogenisation at $1100^{\circ}$C for 5 days and subsequent quenching into water: (a) Backscattered electron image, (b) EBSD map showing the orientation of the c-axis of $\tau$ with respect to the out of plane direction, (c) EBSD grain boundary map showing true twin (blue) with misorientation of the c-axis of $\theta$=75.6$^{\circ}$ and other grain boundaries (black) (The red rectangle shows the region which was used for simulation).}
\label{fig:microstructure_SEM}
\end{figure}

After homogenization, the microstructure of the material was investigated using a Gemini Leo 1530 Scanning Electron Microscope (SEM) and is shown in Figure \ref{fig:microstructure_SEM}(a) in backscattered electron contrast. The microstructure consists of irregularly shaped grains and many planar interfaces, as was observed for C-doped MnAl elsewhere.\cite{Bittner2015a} It should be noted that no indication of the occurrence of phases other than $\tau$ was found using microscopy or with X-ray diffraction measurements.

Electron backscatter diffraction (EBSD) was used to determine the local crystallographic orientation at individual points on the sample. A regular grid of measurement points with a spacing of 500 nm was chosen in the analysis region, yielding in total 40,000 data points. 
The resulting map of easy axis orientation is shown in Figure \ref{fig:microstructure_SEM}(b), represented by the angle from the out-of-plane axis. 
From the EBSD data, the interface distribution was calculated from the misorientation of neighbouring data points. It was found that the so-called ``true'' twins have an angle of misorientation between the magnetically easy, crystallographic c-axis on either side of the twin boundary equal to $\theta$=75.6$^{\circ}$.\cite{Bittner2015a} The spatial distribution of true twin and other grain boundaries is visualized in Fig. \ref{fig:microstructure_SEM}(c). Approximately 36\% of the total interface length is represented by true twins. 

\section{\label{sec:simulation}Simulation}
\subsection{\label{sec:simulation-method}General method}
Micromagnetic simulations were performed by solving the time-dependent Landau-Lifschitz-Gilbert (LLG) equation, using the finite element method (FEM) and using the boundary element method (BEM) to calculate the demagnetizing field.\cite{schrefl2007numerical} 
Room temperature material properties were obtained from the literature with uniaxial magneto-crystalline anisotropy constant $K_{u}$ = 1.7 MJ/m$^{3}$, saturation magnetization $\mu _{0} M_{\mathrm{s}}$ = 0.75 T and exchange stiffness constant $A$ = 2.69 pJ/m.\cite{Rothwarf1981, Coey2010, Ram1983} 
This corresponds to a theoretical anisotropy field of $\mu _{0} H_{a} = 2K_{u}/M_{\mathrm{s}}$ = 5.7 T, an exchange length $L_{\mathrm{ex}} = \sqrt{A/K_{u}} = 1.26$ nm and a theoretical Bloch domain wall width\cite{goll2007micromagnetism} of $\delta = \pi L_{\mathrm{ex}} = 3.95$ nm. Theoretically the highest maximum energy product is $(BH)_{\mathrm{max}} = \mu _{0} M_{s}^{2}/4 = 112$~kJ/m$^{3}$ (14.0 MGOe), providing that the coercive field, $H_{c}$, exceeds half of the remanence $H_{c}\geq M_{r}/2$.\cite{Coey2010} 

Following the work of Rave and co-workers, based on $L_{\mathrm{ex}}$ we use a finite element discretization scheme with minimum edge length 1 nm.\cite{rave1998corners} 
Adaptive meshing is utilized with a fine mesh in areas of the model where we expect domain walls to be formed or located, while a coarser mesh size of up to 3 nm edge length is used elsewhere.  This allows sufficient accuracy for our critical field calculations while curbing the overall model size. The consequences of including such coarsely meshed regions on the interpretation of the simulation results will be discussed in detail in Sections \ref{sec:simulation-APBs} and \ref{sec:simulation-twins}. 

Specific initial magnetization configurations, such as a uniform state or a domain wall state, are created by specifying the magnetization for different regions in the model independently. Then at the start of the simulation remanence is reached in a short time under zero field, whereupon the magnetization configuration is found in a local energy minimum corresponding to the desired state. 

Critical fields, such as the coercive field $H_{c}$, can be estimated from the response of the volume-averaged magnetization $\vec{M}$ to a time-dependent external magnetic field $\vec{H_{ext}}$. A time-varying, spatially-uniform external magnetic field is applied over the whole sample, the strength of which starts at zero for a short time and is then increased linearly at a rate of 0.2 T/ns. Where stated, a small field angle is applied to break the symmetry of the numerical model and hence help to compute realistic critical fields.
Adaptive meshing schemes can introduce artificial domain wall pinning, that makes the usual definition of $H_{c}$ as $H_{\mathrm{ext}}=|\vec{H_{\mathrm{ext}}}|$ when $M/M_{\mathrm{s}}=0$ unsuitable. Critical fields such as domain wall nucleation field, $H_{nuc}$ and domain wall de-pinning field, $H_{depin}$ are estimated accurately despite this by probing the local magnetization at specific locations for a change in polarization of the magnetization component of interest. 
This approach is valid since as soon as a domain wall is released from a pinning potential it is able to move along a smooth nanowire very rapidly relative to the rate of change of the field. However, at smaller $H_{\mathrm{ext}}$ the domain wall velocity is slower, meaning $H_{\mathrm{depin}}$ might be slightly over-estimated. 

Across a magnetically-active APB the magnetic moments are coupled antiferro-magnetically. In contrast to a domain wall the change of the magnetization direction at an APB is discountinuous.\cite{zijlstra_coping_1979} 
In order to model APBs within the framework of the finite element method, we split the mesh at the APB. The exchange energy at the APB is approximated as 
$E = -\left(A^{*}/a\right) \int_S \mathbf{m}_{\mathrm{left}} \cdot \mathbf{m}_{\mathrm{right}} dS$.\cite{suess2009effect} The integral is over the surface area of the APB.
The prefactor in front of the integral is in units of J/m$^2$ and gives the strength of the interfacial coupling. 
The numerator, $A^{*}$, is the exchange stiffness across the interface in units of J/m. 
The denominator, $a$, was set to $0.5$~nm, corresponding to the gap size at the split. The unit vectors of the magnetization at the left and the right hand sides of the interface 
are denoted by 
$\mathbf{m}_{\mathrm{left}}$ and $\mathbf{m}_{\mathrm{right}}$.

\subsection{\label{sec:simulation-APBs}Anti-phase boundaries}

\begin{figure}
\includegraphics[width=1.0\columnwidth]{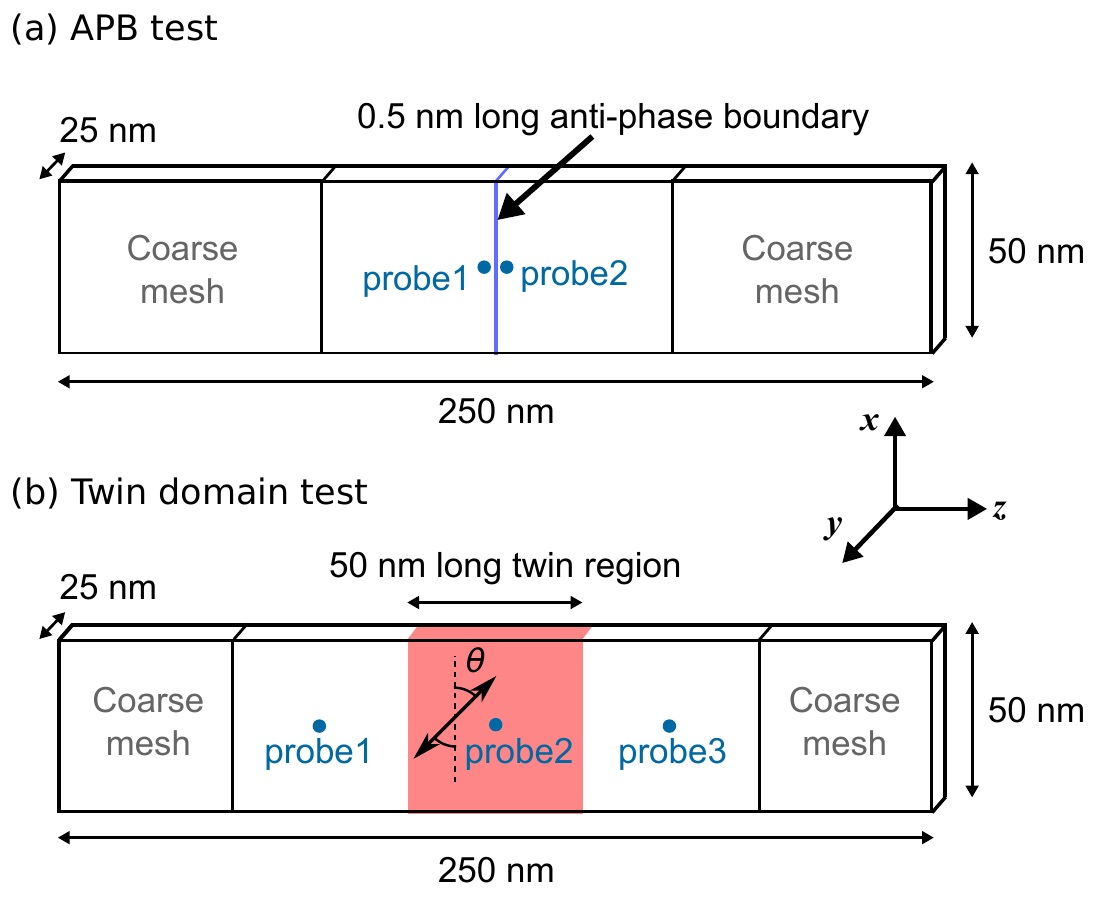}
\caption{Schematics of the micromagnetics simulation models for determining the effects of (a) anti-phase boundaries and (b) twin domains. }
\label{fig:schematics}
\end{figure}

The effects of APBs on the nucleation and pinning of domain walls are assessed using a single-grain cuboidal model of dimensions 50 nm $\times$ 25 nm $\times$ 250 nm (Figure \ref{fig:schematics}a).  
We use $A^{*} = cA$ so that $c$ is an effective coupling constant across the boundary. 
For anti-phase boundaries $c$ is negative, leading to anti-ferromagnetic (AFM) coupling across the boundary. For positive values of $c$ the coupling across the boundary is ferromagnetic (FM). Since experimental estimates of the strength of such coupling are not yet available, we investigate a deliberately wide range of values $-10.0 \leq c \leq 10.0$. A field angle of 5$^{\circ}$ from the easy axis is used. 

Beginning with a domain wall positioned at the boundary, $H_{depin}$ is estimated as the $H_{\mathrm{ext}}$ required to produce a flip in the $M_{x}$ polarity at a probe (``probe2'' in Figure \ref{fig:schematics}a) that is positioned a small distance (1 nm) from the boundary. This indicates that the domain wall has de-pinned from the boundary and propagated through the right hand side of the model, which was initially magnetized in opposition to the external field. 

Likewise, starting with a uniform configuration, $H_{nuc}$ is defined as the field strength required to nucleate a domain wall at the anti-phase boundary. Precise values of $H_{nuc}$ are obtained by probing for a polarity flip of $M_{x}$ a small distance (again 1 nm) either side of the boundary (``probe1'' or ``probe2'' in Figure \ref{fig:schematics}a). Because of symmetry, a domain wall could be formed and propagate in either direction along the sample. 

\begin{figure}
\includegraphics[width=1.0\columnwidth]{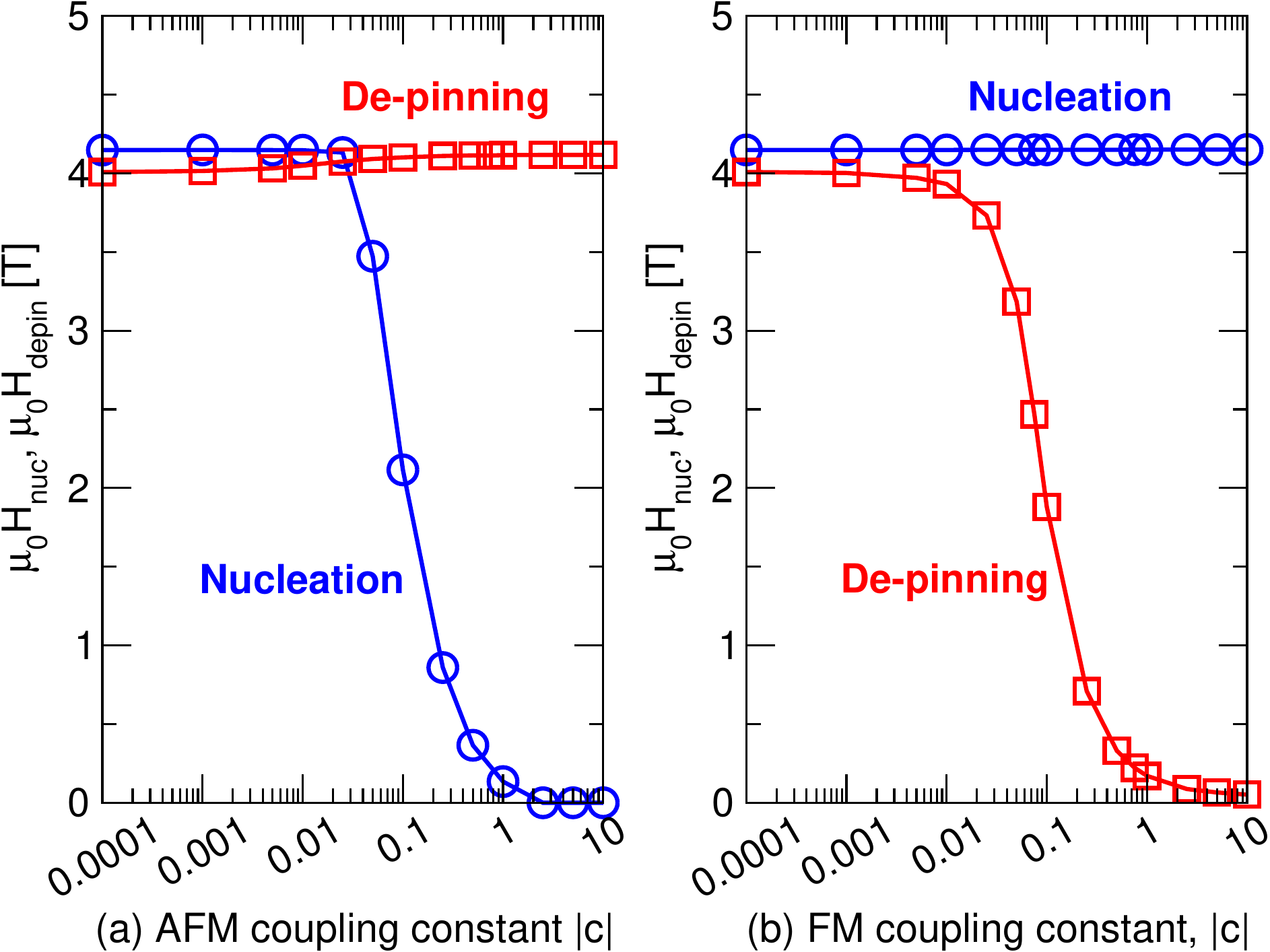}
\caption{Calculated fields required for de-pinning and nucleating a domain wall from an anti-phase boundary, as a function of coupling strength with (a) $c < 0$, corresponding to anti-ferromagnetic (AFM) coupling and (b) $c > 0$, corresponding to ferromagnetic (FM) coupling. }
\label{fig:results-apb}
\end{figure}  

Figure \ref{fig:results-apb} contains computed values of $H_{nuc}$ and $H_{depin}$ for both AFM coupling ($c < 0$) and FM coupling ($c > 0$). 
The main point to note is that the introduction of AFM coupling across the boundary makes it more energetically favorable for a domain wall to be present at the boundary. The larger the AFM coupling becomes (more negative $c$), the lower the required external field becomes to nucleate a domain wall, and the higher the required field becomes to depin a domain wall from the boundary. 
The contrary is true for FM coupling. As the FM coupling at the boundary gets larger the cost in terms of exchange energy of maintaining a domain wall at the boundary becomes far too large and so the field required to depin a domain wall goes down, and the field required to nucleate a domain wall becomes slightly larger. 
Exchange energy effects dominate domain wall behavior since the tight rotation of the local magnetization vector across a domain wall boundary increases the exchange energy contribution to the total Gibbs free energy of the system. 

\begin{figure}
\includegraphics[width=0.9\columnwidth]{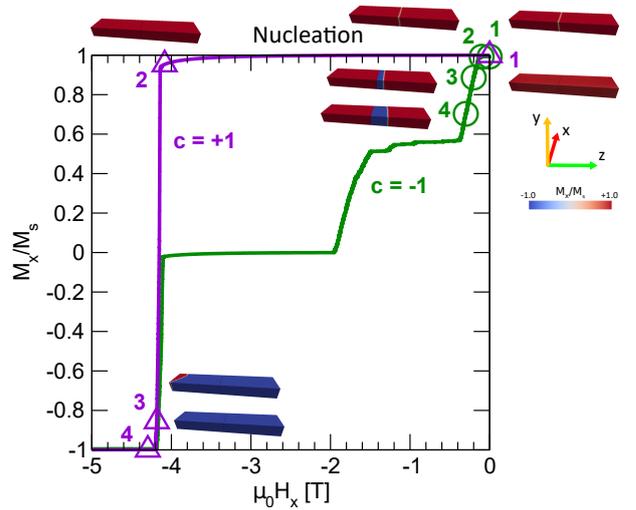}
\caption{$M$-$H$ reversal curves for domain wall nucleation at an APB with FM ($c=+1.0$) and AFM ($c=-1.0$) coupling. Inset images show the magnetization configuration of the samples corresponding to various stages during the reversal process. }
\label{fig:apb_comp_nuc}
\end{figure}

\begin{figure}
\includegraphics[width=0.9\columnwidth]{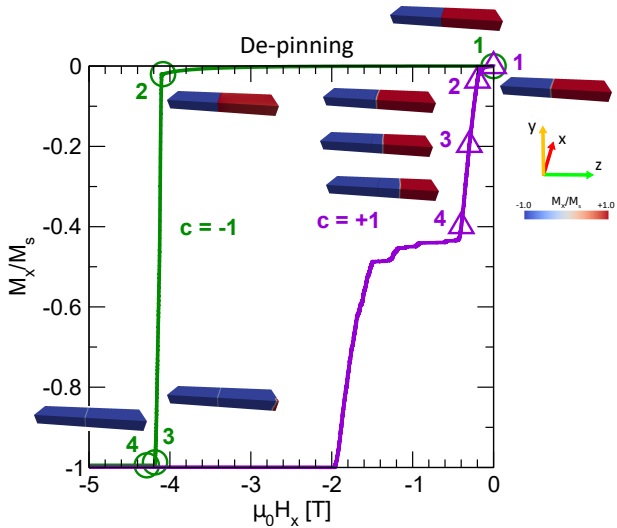}
\caption{$M$-$H$ reversal curves for domain wall de-pinning from an APB with FM ($c=+1.0$) and AFM ($c=-1.0$) coupling. Inset images show the magnetization configuration of the samples corresponding to various stages during the reversal process. }
\label{fig:apb_comp_depin}
\end{figure}

Examples of the hysteresis curves for the usual FM ($c=+1.0$) coupling and AFM ($c=-1.0$) coupling are given for both nucleation (Figure \ref{fig:apb_comp_nuc}) and de-pinning (Figure \ref{fig:apb_comp_depin}). 

For nucleation (Figure \ref{fig:apb_comp_nuc}) with $c=+1$, starting with the remanent state (label 1) we do not see any change until $H_{\mathrm{ext}} \approx 4.1$~T (label 2), at which point the Zeeman energy contribution is high enough that a DW state becomes energetically favorable. In order to create the reversed domain, a second, regular domain wall must also be created and propagate through the sample. At $c=+1.0$ the boundary provides no pinning potential, so immediately after nucleation the DW rapidly propagates through, reversing the whole sample (labels 3-4). 
For $c=-1$, however, the AFM coupling in the initial uniform configuration (label 1) introduces a very high exchange energy locally at the boundary. A DW configuration is already energetically favorable and it takes only small $H_{\mathrm{ext}}$ to overcome the energy barrier associated with introducing a domain wall into the system (label 2). A second domain wall on the opposite side of the reversal domain then propagates through the sample (labels 2-4) before being subject to an artificial pinning potential at the edge of the coarsely-meshed end region of the model. In this case the mesh pinning potential is higher than the current $H_{\mathrm{ext}}$ so reversal is hindered until $H_{\mathrm{ext}}$ increases to around 2.5 T, whereupon the domain wall is expelled. We reach a half-reversed state where the volume-averaged magnetization is zero. At 4.1 T another regular domain wall is nucleated at the APB and propagates in the opposite direction until the sample is fully reversed. 

Onto de-pinning (Figure \ref{fig:apb_comp_depin}), in the $c=+1$ case we start with an existing DW at the APB (label 1). Since the local properties match that of bulk, there is no pinning potential. The DW depins (label 2) and starts to propagate through the sample (label 3) at very small $H_{\mathrm{ext}}$. It then meets the coarse mesh region (label 4) and is subject to artificial pinning potentials that are only overcome at around 1.6 T, at which point full reversal is achieved. 
In the case of $c=-1$ the AFM coupling leads to reduced DW energy at the APB (label 1), so a much larger $H_{\mathrm{ext}} \approx 4.1$~T is required to depin (label 2). Reversal is fast and the domain wall is expelled rapidly (labels 3-4). 

In cases where the artificial mesh pinning is not observed, $H_{\mathrm{ext}}$ is already larger than the pinning potential generated by the coarse mesh region. 
The artificial mesh pinning we have demonstrated does not affect our estimations of $H_{nuc}$ and $H_{depin}$, because in both cases we only need to know the fields associated with the initial nucleation and de-pinning events. Further information from later stages of the simulation is not needed. We include the full reversal curve data here for completeness and to demonstrate that care should be taken when interpreting results from models where adaptive meshing has been utilized, particularly at lower critical field strengths comparable to the artificial mesh boundary de-pinning field. 

Although the true strength of coupling in APBs is not known from experiment, our simulations show that for values even one tenth the size of the usual ferromagnetic exchange coupling in MnAl, 
the nucleation field is already halved and the de-pinning field is doubled. 
For reasonable estimates of $c$, nucleation at APBs will likely occur throughout the sample. 
We see a strong inclination for domain walls to nucleate and become strongly pinned at the APBs. This result is in good accordance with Lorentz TEM data from the literature, which showed that APBs are always decorated with domain walls.\cite{Jakubovics1977b, VanLanduyt1978a, Jakubovics1978b, Houseman1983d} 
Hot extruded MnAl magnets have superior magnetic properties compared to as-transformed MnAl and it has been proposed that the lack of APBs in the hot deformed state is one major contributing factor to this.\cite{VanLanduyt1978a}

%

\subsection{\label{sec:simulation-twins}Twin domain boundaries}

In a separate set of simulations, the influence of twin boundaries is calculated using a cuboidal model of size 50 nm $\times$ 25 nm $\times$ 250 nm, containing a 50 nm long twin domain region at the center (Figure \ref{fig:schematics}b). This twin domain is defined by rotation of the uniaxial anisotropy axis relative to the rest of the sample. In the end regions the uniaxial anisotropy angle is parallel to the $x$ axis, whereas in the twin region the crystallographic orientation is tilted by a certain in-plane twinning angle in the range $-180^{\circ} \leq \theta \leq 180^{\circ}$. 
$H_{nuc}$ is defined as the $H_{\mathrm{ext}}$ required to flip the polarization at either ``probe1'' or ``probe3'' in Figure \ref{fig:schematics}, both 25 nm from a boundary, indicating that a domain wall configuration has occurred. 
Similarly, $H_{\mathrm{depin}}$ is defined as the field required to depin a domain wall from the twin region. The situation is complicated by the possibility of beginning with a domain wall at either the left-hand (L) or right-hand (R) ends of the twin region, so we define $H_{\mathrm{depin,L}}$ as the field required to move a DW beginning at the left-hand boundary past the central probe (``probe2'') and $H_{\mathrm{depin,R}}$ as the field required to move a DW beginning at the right-hand boundary past the right hand probe (``probe3''). 
No additional field angle is applied so that $\theta$ also corresponds to the relative angle between the twin easy axis and the applied field. 

\begin{figure}
\includegraphics[width=1.0\columnwidth]{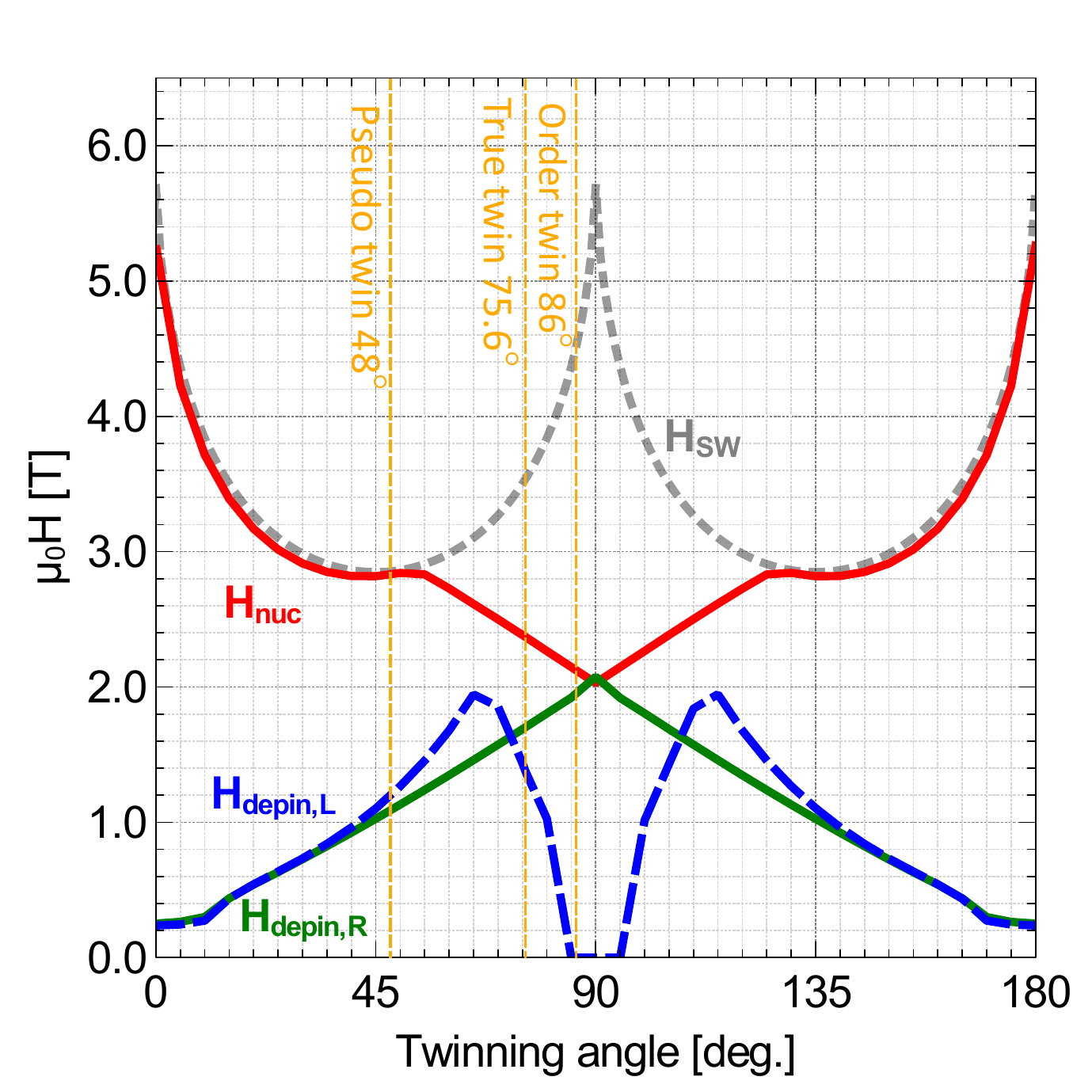}
\caption{Twin domain boundary nucleation field $H_{\mathrm{nuc}}$ and de-pinning fields for left $H_{\mathrm{depin,L}}$ and right $H_{\mathrm{depin,R}}$ initial positions as a function of twinning angle $\theta$. The Stoner-Wohlfarth field $H_{\mathrm{SW}}$ is shown for comparison. The dotted line indicates the experimentally-determined twinning angles of the three known types of twin, the pseudo, true and order twins with angles 48$^\circ$, 86$^\circ$, and 75.6$^\circ$, respectively. The smallest nucleation field is found for the order twin which in turn shows the highest pinning field. On the other hand, the pseudo twin shows a high nucleation field but a small pinning field.}
\label{fig:results-twins}
\end{figure}

The nucleation field $H_{\mathrm{nuc}}$ and de-pinning field $H_{\mathrm{depin}}$ for domain walls at twin boundaries in our single-grain model are given as functions of twinning angle $\theta$ in Figure \ref{fig:results-twins}. Only angles in the range $0^{\circ} \leq \theta \leq 180^{\circ}$ are shown in the plot as the calculated values are again exactly symmetrical around the vertical axis $\theta=0$. Further, all of the data series are symmetrical around $90^\circ$. 
These symmetries appear due to the uniaxial nature of the magnetic materials and indicate that the primary influence is the relative magnetization angle between neighboring regions.  
The twinning angles for the three experimentally-determined twin types, the pseudo, true and order twins with angles 48$^\circ$, 86$^\circ$, and 75.6$^\circ$, respectively, are indicated. 

\begin{figure}
\includegraphics[width=1.0\columnwidth]{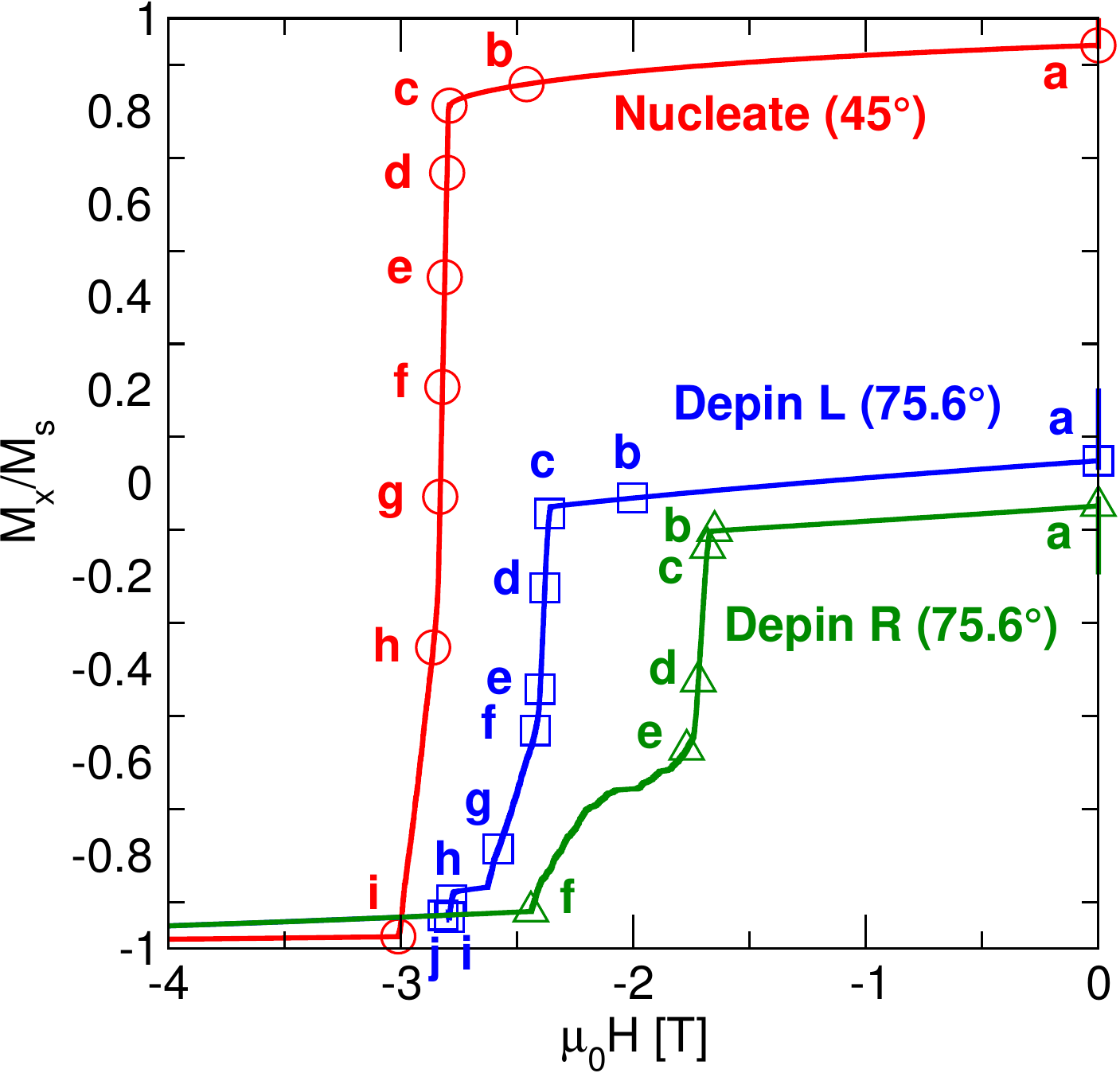}
\caption{$M$-$H$ curves for the nucleation mechanism with $\theta=45^{\circ}$ and for de-pinning mechanisms at both starting position with $\theta=75.6^{\circ}$. The labels a, b, c, \emph{etc.} correspond to those in Figure \ref{fig:TwinsImages}}
\label{fig:TwinsHysteresis}
\end{figure}

\begin{figure}
\includegraphics[width=1.0\columnwidth]{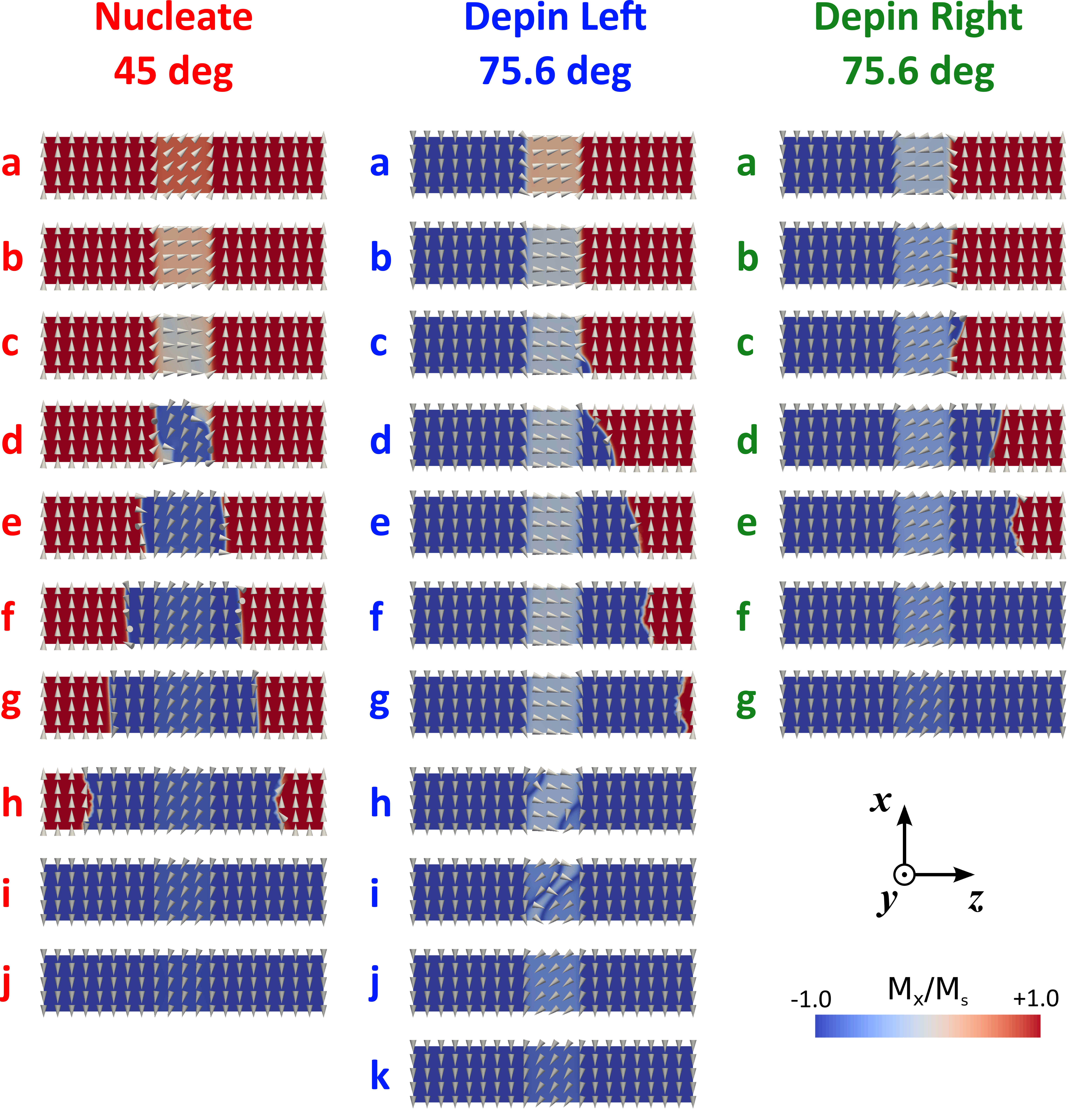}
\caption{Magnetization configurations for nucleation and de-pinning processes with specific twinning angles. The labels a, b, c, \emph{etc.} correspond to those in Figure \ref{fig:TwinsHysteresis}}
\label{fig:TwinsImages}
\end{figure}

$M$-$H$ curves for the nucleation mechanism with $\theta=45^{\circ}$, and for de-pinning mechanisms at both starting position with $\theta=75.6^{\circ}$, are given in Figure \ref{fig:TwinsHysteresis}. The magnetization configurations  corresponding to the labels a, b, c, \emph{etc}. are visualized in Figure \ref{fig:TwinsImages}.

With a twinning angle of $\theta=0^\circ$ we have the baseline situation of a uniform sample with homogeneous properties, where $H_{\mathrm{nuc}}$ approaches the theoretical anisotropy field $H_{a}=2K_{u}/\mu_{0}M_{s}$. In this case the nucleation of a domain wall is energetically unfavorable and requires a strong field. 
As the twinning angle is increased the nucleation field matches closely the Stoner-Wohlfarth field $H_{\mathrm{SW}}$, which describes small, homogeneous particles.\cite{SW}  

\begin{equation}
	H_{\mathrm{SW}} = H_{a}(\mathrm{cos}^{2/3}\theta + \mathrm{sin}^{2/3}\theta)^{-3/2}
\end{equation}

\begin{figure}
\includegraphics[width=1.0\columnwidth]{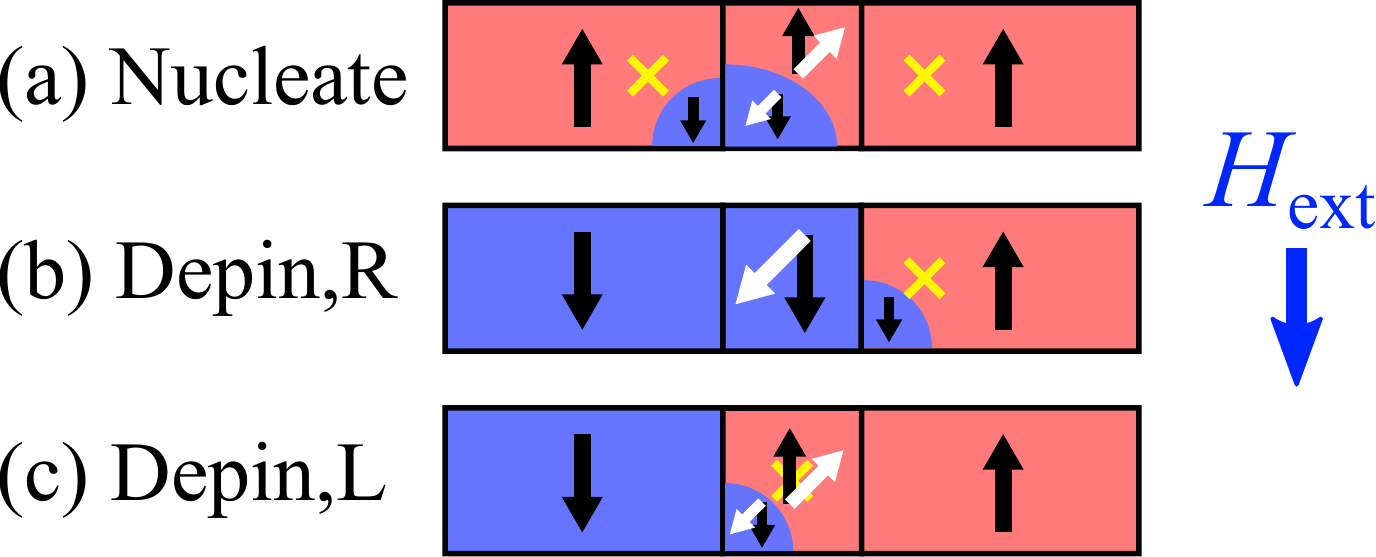}
\caption{Schematics describing the mechanisms involved with the prescribed events of (a) domain wall nucleation, (b) domain wall de-pinning from the right hand (``R'') twin boundary and (c) domain wall de-pinning from the left hand (``L'') twin boundary. The probes used in each case are indicated by a yellow cross. Black arrows indicate the magnetization. White arrows indicate the magnetization as modified by the change in easy axis imparted by a non-zero twinning angle, which lead to a change in the relative angle between the easy axis and the external field. }
\label{fig:twins-explanation}
\end{figure}

The increasing relative angle between the easy axis of the twin region and the applied field increases the magnetic torque effect, which is encapsulated in the Landau-Lifshitz-Gilbert (LLG) equation of micromagnetic motion. 
In order to nucleate a domain wall at the twin region we actually have to reverse one half of the sample, which involves de-pinning and propagating another DW through the main phase (Figure \ref{fig:twins-explanation}a). The initial creation of the domain wall involves the nucleation of a small reversal volume, through localized rotation, that is known to be dependent on the field angle.\cite{bance2014micromagnetics, bance2014influence, bance2014hard, bance2014grain, bance2015thermal} 
At lower values of $\theta$, $H_{\mathrm{nuc}}$ is always larger than the de-pinning field, so the critical field follows the S-W curve. 
However, at larger $\theta$ the domain wall configuration state becomes energetically so favorable over the uniform state that a domain wall is formed at the left hand boundary through de-pinning, pre-empting the reversal of the twin region. The higher the $\theta$, the lower the field at which this occurs, until at $\theta=90^{\circ}$ we converge with the $H_{\mathrm{depin,R}}$ value. 
$H_{\mathrm{nuc}}$ is now simply the field required to depin and so we find a bottleneck; $H_{\mathrm{nuc}}$ can never be smaller than $H_{\mathrm{depin,R}}$. 


Onto the de-pinning fields, we begin at $\theta=0^\circ$ with very low field values required, for both initial configurations, since this corresponds to a normal, uniform sample. These fields are not quite zero, since at low field strengths it takes some finite time for the domain wall to move past our probe positions. 
$H_{\mathrm{depin,R}}$ is effectively the field required to depin and propagate a DW through the main phase to the right (that has the usual micromagnetic parameters), overcoming only the pinning potential at the right hand boundary, as described in Figure \ref{fig:twins-explanation}b. This pinning potential varies linearly with the twinning angle; as the angle increases so does $H_{\mathrm{depin,R}}$, as the exchange energy associated with a domain wall at the boundary drops and the energy barrier associated with de-pinning and introducing a 180$^{\circ}$ domain wall into the main phase increases. 
At a maximum value of $90^\circ$ it matches $H_{\mathrm{nuc}}$ with around 2.05 T, and then drops above $90^\circ$ due to the uniaxial symmetry. 
Since a domain wall that becomes de-pinned moves into the main phase, we do not see any influence of twinning angle on the result, as field angle relative to the main phase easy axis is constant. 

In the case of $H_{\mathrm{depin,L}}$ we also need to overcome a pinning potential that increases with twinning angle, so a similar initial increase is observed. However, in this case the domain wall must enter the twin domain and the relative angle between the external field and the easy axis is also changing, as described in Figure \ref{fig:twins-explanation}c. The Stoner-Wohlfarth minimum is again noticed at $45^\circ$, followed by an increase in $H_{\mathrm{depin,L}}$ at higher angles. 
At higher angles yet the central region is able to flip polarity by in-plane rotation, which happens at a much lower field, so $H_{\mathrm{depin,L}}$ drops rapidly. Interestingly, for such angles the central twin domain does not fully reverse until after the domain wall has already de-pinned (Figure \ref{fig:TwinsImages}, label h), since the in-plane rotation must overcome an energy barrier corresponding to being oriented perpendicular to the easy axis. 
Finally, at $90^\circ$ no field is required since the domain wall effectively spans the entire twin domain. 

The field required to fully depin a DW from such a boundary clearly depends not only on the twin angle, but also on the initial DW position, according to some deceptively complex reversal mechanisms. 
For each of the three known twin angles, the calculated nucleation and de-pinning fields are considerably lower than $H_{a}$, so the presence of twin defects should be expected to degrade performance of such a permanent magnet. 
The type of twin defect that we expect to be most detrimental to such a permanent magnet is the 86$^{\circ}$ order twin, since it corresponds to the lowest nucleation field and the highest de-pinning field of all the twin types. Easier nucleation and stronger pinning of magnetic domain walls at the twin boundaries reduces remanent magnetization and coercivity, which in turn lowers the $(BH)_{\mathrm{max}}$. The second biggest impact is from the 75.6$^{\circ}$ true twins, following similar arguments.  

\subsection{\label{sec:simulation-granular}Granular Structure}

In an effort to extend the investigation to MnAl magnets with realistic granular structures, a technique was developed to convert EBSD data to finite element models. The EBSD scan produces a regular gridded 2D map of crystallographic orientation over the sample. This preserves the grain structure of the sample, including any twin boundaries. The sample used in this report is here visualized to show both c-axes angle (Figure \ref{fig:microstructure_SEM}b) and the grain boundaries (Figure \ref{fig:microstructure_SEM}c). 
This gridded data is processed using the mesh generation and processing toolbox iso2mesh,\cite{Fang2009} first converting individual grains into regions and then extruding them into prismatic volumes. The resulting thin-film geometry is meshed using tetrahedral finite elements. 

We face some challenges with this process. 
The grains are typically tens of microns in diameter but, as stated earlier, we should obey certain minimum finite-element mesh-size requirements that are dependent on the materials' intrinsic properties.\cite{rave1998corners} A full-sized model would therefore contain many elements and a solution becomes prohibitively expensive to compute, so we are forced to model a cropped region of the full sample instead. However, if we crop to a size that is manageable (in this case around 1 micron grain diameter) the wider texture of the sample is no longer visible. An additional problem is that when we zoom into a region approaching the finite scan resolution size (0.5 $\mu$m) the grain texture is no longer visible. 
A compromise is reached by scaling certain cropped regions that still contain sufficient grain structure. 
With this approach we can approximate the effects of a realistic granular structure; if the grains are still above the typical coherence size then the reversal mechanism for each grain should be similar to those of the original sample.\cite{Coey2010}


Despite some unidentified errors from the meshing algorithm, two models with scaling ratios of 1:500 and 1:50 were successfully generated, where the feature sizes are 500x and 50x smaller than reality. The models contain an additional region that is already magnetically reversed at the start of the simulation to allow a consistent initial domain wall configuration. 
For the 500x model (marked by the red rectangle in Figure \ref{fig:microstructure_SEM}c) the dimensions are 45 nm $\times$ 35 nm $\times$ 3 nm with a mesh size of 0.5 nm necessary to resolve some of the smaller features. 
For the 50x model a smaller region is cropped, resulting in a model size of 190 nm $\times$ 150 nm $\times$ 10 nm, with a finite element mesh size of 1 nm. 
The 500x model is further from the true sample size but preserves more of the original domain texture. By using two scaling sizes we can make broad predictions, extrapolating towards the performance of a true-to-size sample. 

\begin{figure}
\includegraphics[width=1.0\columnwidth]{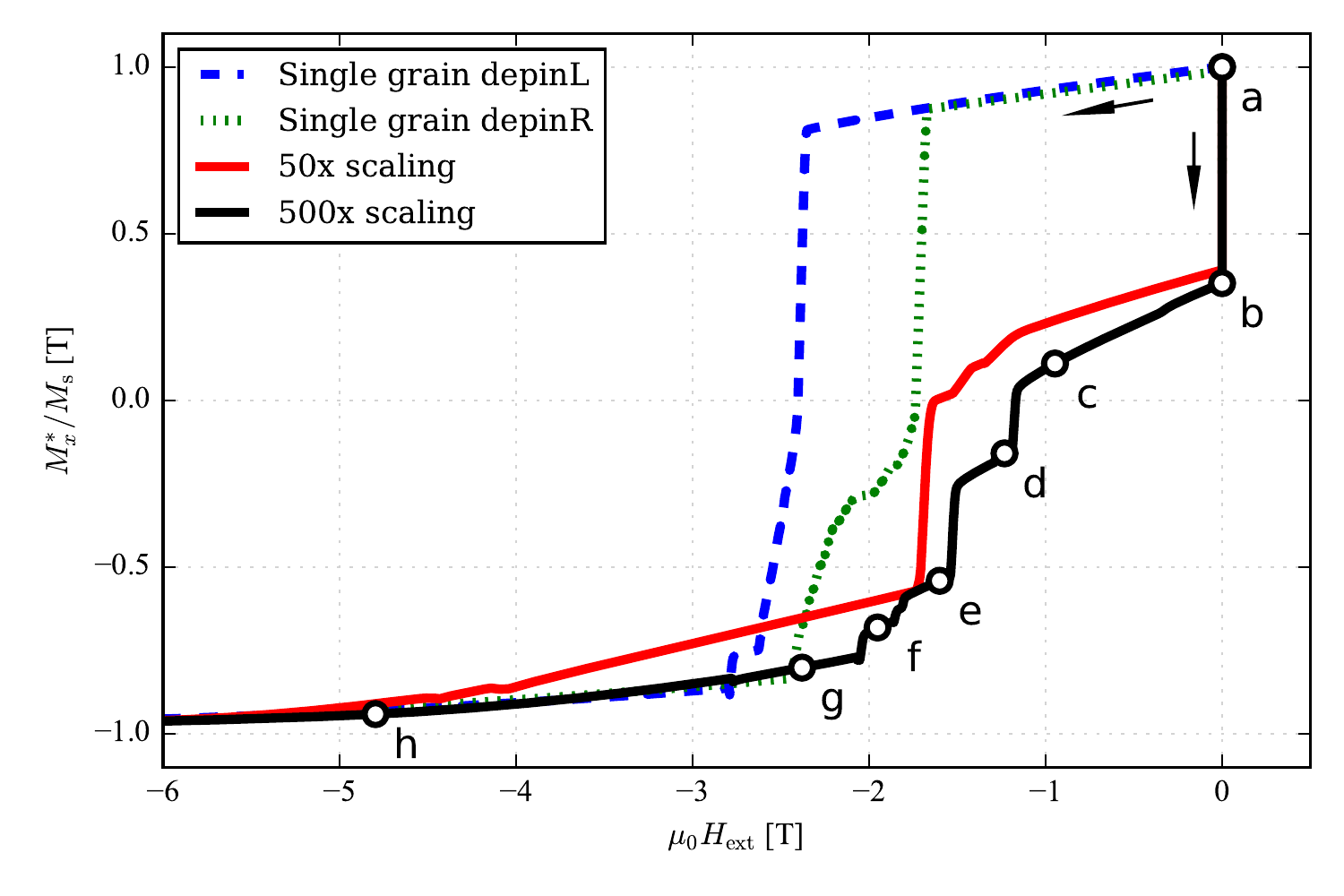}
\caption{Reversal curves for $\tau$-MnAl models of realistic grain and domain texture and various size scaling factor. The curves for the single grain twin domain models are shown for direct comparison. As we approach a model containing many grains the squareness is drastically reduced.}
\label{fig:results-granular}
\end{figure}

$M$-$H$ curves from the reversal process are calculated by saturating the sample along the +x direction, allowing the sample to relax to its remanent magnetization state and then applying an opposing external field whose magnitude starts at zero and increases by 0.2 T/ns. The curves for both models are compared to that of the single grain model with a true twin angle of $75.6^{\circ}$ (Figure \ref{fig:results-granular}), for both left and right initial positions. 
These magnetization curves are adjusted to consider only the volumes that were originally opposing the external field. This allows us to easily compare the coercive field and calculate values for $(BH)_{\mathrm{max}}$. 
When we move from the single grain models to the granular models there is a marked reduction in squareness of the reversal curves. This reduction is even more pronounced for the 500x where the granular texture is better reproduced. Also there is a large reduction in remanent magnetization as the magnetization follows the local magneto-crystalline anisotropy of the individual grains. 
The coercive field for the more granular sample is also lower. This suggests that a true-to-scale version might have lower coercivity still. 

\begin{figure}
\includegraphics[width=1.0\columnwidth]{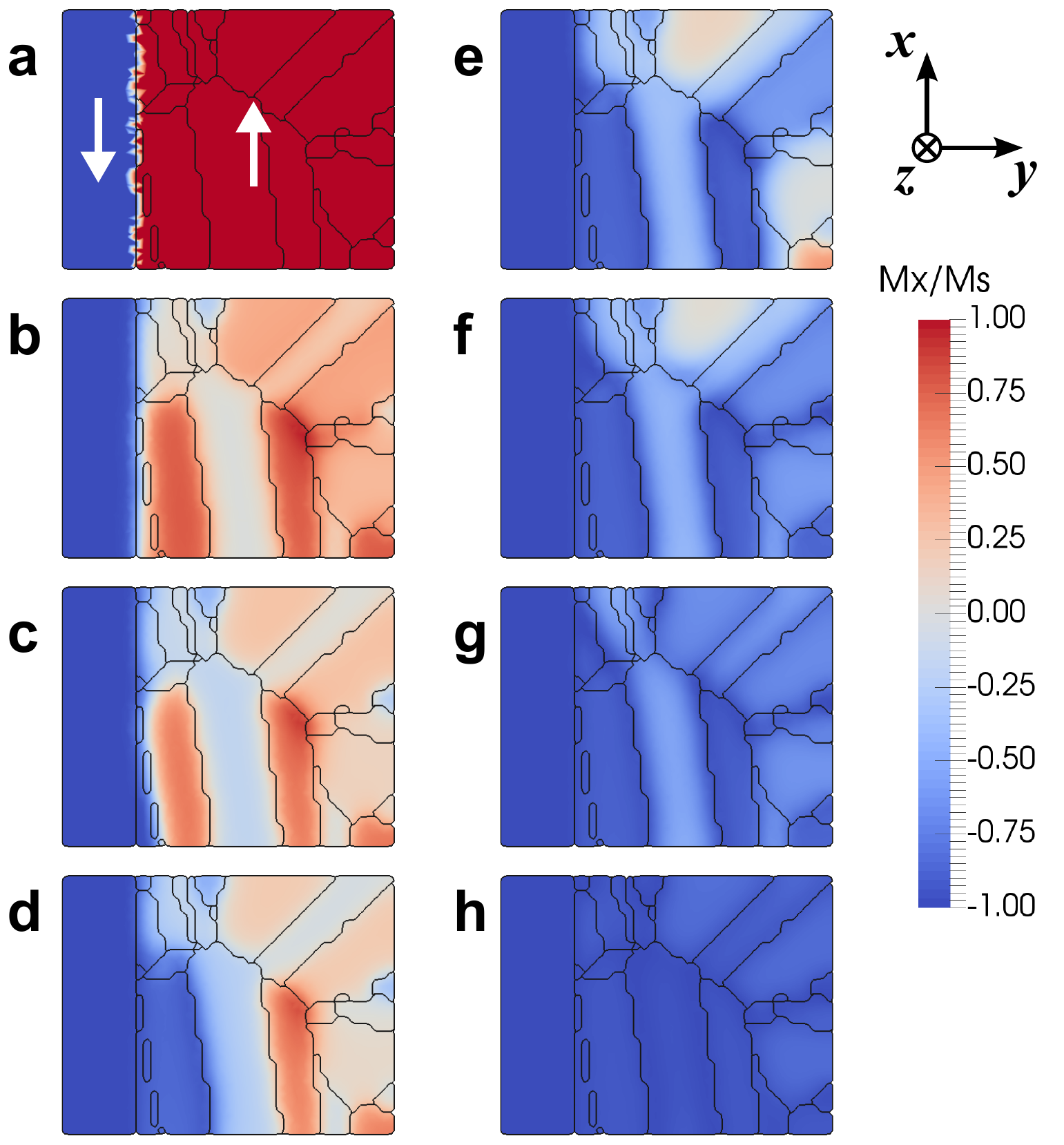}
\caption{Stages of reversal in granular model with 500x scaling.  }
\label{fig:results-granularStages}
\end{figure}

Visualizations of the 500x model reversal process are given in Figure \ref{fig:results-granularStages}, corresponding to the matching labels on the reversal curve of (Figure \ref{fig:results-granular}). The initial state (a) is saturated with two opposing regions in order to begin with a domain wall in the sample. This relaxes to remanence (b) under zero field. As the external field strength increases through (c)-(g) individual regions undergo reversal either by rotation or by propagation of domain walls through the structure, until (h) where the sample is approaching saturation. 
At the remanent state (b) some small grains close to the edge of the model are already reversed. This effect is similar to domain formation in soft magnets, resulting from high local demagnetization energy, especially important where there might be surface charges at the boundaries of our finite-sized model. 

\begin{figure}
\includegraphics[width=1.0\columnwidth]{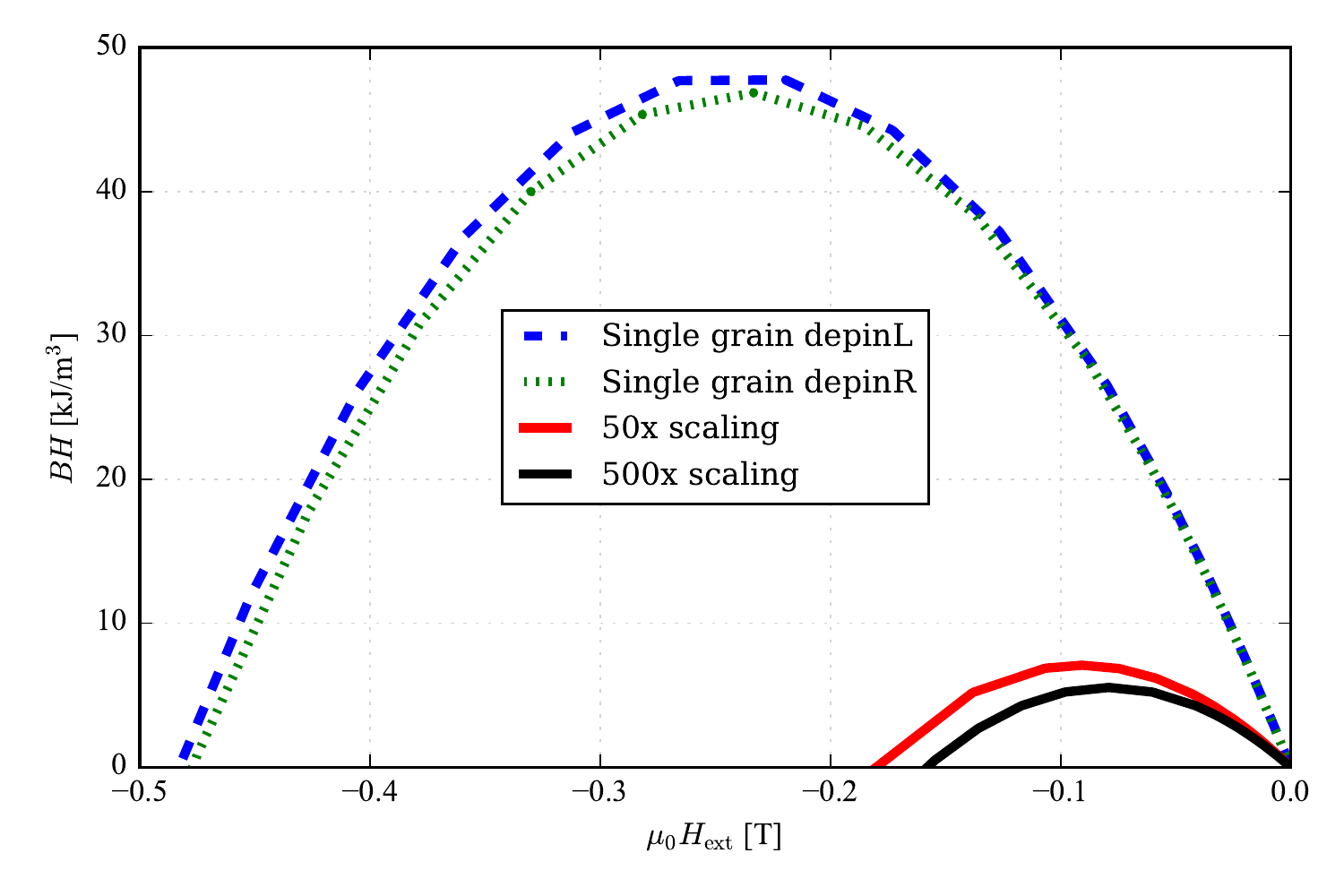}
\caption{Reversal curves for $\tau$-MnAl models of realistic grain and domain texture containing twin defects. The curves for the single grain twin domain models are shown for direct comparison. As we approach a model containing many grains the squareness is drastically reduced.}
\label{fig:results-granular-BH-H}
\end{figure}

Converting in the usual way, $BH$ is plotted against $H$ in Figure \ref{fig:results-granular-BH-H}. $(BH)_{\mathrm{max}}$ values for the 50x and 500x scaled granular samples are calculated to be 7.1 kJ/m$^{3}$ (0.89 MGOe)and 5.5 kJ/m$^{3}$ (0.69 MGOe), respectively. These are well short of the theoretical maximum of 112~kJ/m$^{3}$ (14.0 MGOe), a consequence of their reduced remanence, coercivity and extremely poor squareness of the reversal curves. 
This agrees well with the standard knowledge that MnAl permanent magnets with twin-containing texture have much lower performance than those that have been further processed.
For the single grain models, $(BH)_{\mathrm{max}}$ is calculated to be 47.8 kJ/m$^{3}$ (6.0 MGOe)and 46.9 kJ/m$^{3}$ (5.9 MGOe), reinforcing the suggestion that if grains are crystallo-graphically better-aligned then $(BH)_{\mathrm{max}}$ is improved. 

We did not simulate the presence of APBs in this stucture. However, it would be reasonable to assume that the addition of APBs should further decrease the $(BH)_{\mathrm{max}}$ by increasing the number of domain wall nucleation sites. Since the AFM coupling at APBs would probably be accompanied by very low nucleation fields, the remanent magnetization should be further reduced. 

According to our work, twin domain boundaries act as nucleation and pinning sites for magnetic domain walls. However, the nucleation field of twin boundaries are always greater than the pinning field $|H_{\mathrm{nuc}}| \ge |H_{\mathrm{depin}}|$.
For a homogeneous distribution of such boundaries, nucleation will happen throughout the $\tau$-MnAl phase, 
degrading the magnetization of the permanent magnet away from the square-shaped hysteresis behavior 
and reducing the coercivity. This results are consistent with the observation that large grains with a high twin density have a much lower reversal field than do fine grains with a smaller number of twins.\cite{Thielsch2017}

\section{\label{sec:conclusion}Conclusions}

We have developed understanding of the complex interactions of APBs and twins with domain walls in as-transformed $\tau$-MnAl by carrying out micromagnetic simulations. 
The presence of both types of defect was found to encourage domain wall nucleation and pinning. 
Of the three crystallographically different twin-like defects that are found in $\tau$-MnAl, the smallest nucleation field is found for the order twin which in turn shows the highest pinning field. On the other hand, the pseudo twin shows
a high nucleation field but a small pinning field. 
The most common twin type, the true twin, sits somewhere in-between, reducing the domain wall nucleation field and increasing the domain wall de-pinning field relative to the main phase of the material. 
The simulations confirm experimental observations that samples with many twins have a much lower coercivity. 
Models based closely on the results of microstructural characterization have shown that granular structure containing fine twinning drastically reduces the hysteresis loop squareness.  
This lowers the maximum energy product, $(BH)_{\mathrm{max}}$ to just 5 percent of the theoretical maximum, which confirms that one of the main reasons why further processing can improve $(BH)_{\mathrm{max}}$ is due to the removal of twin defects from the microstructure.

\section*{Acknowledgements}
The authors would like to acknowledge funding support from the EU Seventh Framework Program (FP7) project ROMEO: Replacement and Original Magnet Engineering Options 
(Project ID: 309729) and the  EU H2020 Program project NOVAMAG: Novel, critical materials free, high anisotropy phases for permanent magnets, by design (Project ID: 686056).

\section*{References}

\bibliography{actaArXiV}

\end{document}